\documentstyle[12pt,indent,epsf,eqsection,subeqnarray]{article}

\begin{document}

\bibliographystyle{plain}


\title{The 3-State Potts Antiferromagnet on the Hexagonal Lattice}  
\author{
  {\small Jes\'us Salas}                  \\[-0.2cm]
  {\small\it Departamento de F\'{\i}sica de la Materia Condensada and} \\[-0.2cm]
  {\small\it Departamento de F\'{\i}sica Te\'orica} \\[-0.2cm]
  {\small\it Facultad de Ciencias}        \\[-0.2cm]
  {\small\it Universidad de Zaragoza}     \\[-0.2cm]
  {\small\it 50009 Zaragoza SPAIN}       \\[-0.2cm]
  {\small\tt JESUS@JUPITER.UNIZAR.ES}     \\[-0.2cm]
  {\protect\makebox[5in]{\quad}}  
  \\
}
\vspace{0.5cm}

\maketitle
\thispagestyle{empty}   

\def\spose#1{\hbox to 0pt{#1\hss}}
\def\ltapprox{\mathrel{\spose{\lower 3pt\hbox{$\mathchar"218$}}
 \raise 2.0pt\hbox{$\mathchar"13C$}}}
\def\gtapprox{\mathrel{\spose{\lower 3pt\hbox{$\mathchar"218$}}
 \raise 2.0pt\hbox{$\mathchar"13E$}}}
\def\inapprox{\mathrel{\spose{\lower 3pt\hbox{$\mathchar"218$}}
 \raise 2.0pt\hbox{$\mathchar"232$}}}

\vspace{1cm}

\begin{abstract}
We study the 3-state hexagonal-lattice Potts antiferromagnet
by a Monte Carlo simulation
using the Wang-Swendsen-Koteck\'y cluster algorithm.
We study the staggered susceptibility and the correlation length, and 
we confirm that this model is disordered at all temperatures $T\geq0$.  
We also measure the ground-state entropy density. 
\end{abstract}


\clearpage

\newcommand{\be}{\begin{equation}}
\newcommand{\ee}{\end{equation}}
\newcommand{\<}{\langle}
\renewcommand{\>}{\rangle}
\newcommand{\para}{\|}
\renewcommand{\perp}{\bot}

\def\smfrac#1#2{{\textstyle\frac{#1}{#2}}}
\def\half{ {{1 \over 2 }}}
\def\smhalf{ {\smfrac{1}{2}} }
\def\scra{{\cal A}}
\def\scrc{{\cal C}}
\def\scrd{{\cal D}}
\def\scre{{\cal E}}
\def\scrf{{\cal F}}
\def\scrh{{\cal H}}
\def\scri{{\cal I}}
\def\scrk{{\cal K}}
\def\scrm{{\cal M}}
\newcommand{\scrmvec}{\vec{\cal M}_V}
\def\scrmtens{{\stackrel{\leftrightarrow}{\cal M}_T}}
\def\scro{{\cal O}}
\def\scrp{{\cal P}}
\def\scrr{{\cal R}}
\def\scrs{{\cal S}}
\def\ttens{{\stackrel{\leftrightarrow}{T}}}
\def\scrv{{\cal V}}
\def\scrw{{\cal W}}
\def\scry{{\cal Y}}
\def\scrz{{\cal Z}}
\def\tauss{\tau_{int,\,\scrm^2}}
\def\taux{\tau_{int,\,{\cal M}^2}}
\newcommand{\taum}{\tau_{int,\,\vec{\cal M}}}
\def\taue{\tau_{int,\,{\cal E}}}
\newcommand{\imag}{\mathop{\rm Im}\nolimits}
\newcommand{\real}{\mathop{\rm Re}\nolimits}
\newcommand{\tr}{\mathop{\rm tr}\nolimits}
\newcommand{\sgn}{\mathop{\rm sgn}\nolimits}
\newcommand{\codim}{\mathop{\rm codim}\nolimits}
\newcommand{\rank}{\mathop{\rm rank}\nolimits}
\newcommand{\sech}{\mathop{\rm sech}\nolimits}
\def\textprime{{${}^\prime$}}
\newcommand{\longto}{\longrightarrow}
\def\var{ \hbox{var} }
\newcommand{\gtilde}{ {\widetilde{G}} }
\newcommand{\USp}{ \hbox{\it USp} }
\newcommand{\CP}{ \hbox{\it CP\/} }
\newcommand{\QP}{ \hbox{\it QP\/} }
\def\hboxscript#1{ {\hbox{\scriptsize\rm #1}} }

\newcommand{\plotdot}{\makebox(0,0){$\bullet$}}
\newcommand{\plotsmalldot}{\makebox(0,0){{\footnotesize $\bullet$}}}

\def\bsigma{\mbox{\protect\boldmath $\sigma$}}
\def\bpi{\mbox{\protect\boldmath $\pi$}}
\def\btau{\mbox{\protect\boldmath $\tau$}}
\def\bn{{\bf n}}
\def\br{{\bf r}}
\def\bz{{\bf z}}
\def\bh{\mbox{\protect\boldmath $h$}}

\def\betatilde{ {\widetilde{\beta}} }
\def\hatp{\hat p}
\def\hatl{\hat l}

\def\msbar{ {\overline{\hbox{\scriptsize MS}}} }
\def\normalmsbar{ {\overline{\hbox{\normalsize MS}}} }

\def\eff{ {\hbox{\scriptsize\em eff}} }

\newcommand{\reff}[1]{(\ref{#1})}

\newcommand{\Z}{{\bf Z}}
\newcommand{\zed}{{\bf Z}}
\newcommand{\R}{{\bf R}}
\font\srm=cmr7 		
\def\szed{\hbox{\srm Z\kern-.45em\hbox{\srm Z}}}
\def\sR{\hbox{{\srm I}\kern-.2em\hbox{\srm R}}}
\def\C{{\bf C}}



\newtheorem{theorem}{Theorem}[section]
\newtheorem{corollary}[theorem]{Corollary}
\newtheorem{lemma}[theorem]{Lemma}
\newtheorem{conjecture}[theorem]{Conjecture}
\newtheorem{definition}[theorem]{Definition}
\def\proof{\bigskip\par\noindent{\sc Proof.\ }}
\def\qed{\hbox{\hskip 6pt\vrule width6pt height7pt depth1pt \hskip1pt}\bigskip}

%
%
\newenvironment{sarray}{
          \textfont0=\scriptfont0
          \scriptfont0=\scriptscriptfont0
          \textfont1=\scriptfont1
          \scriptfont1=\scriptscriptfont1
          \textfont2=\scriptfont2
          \scriptfont2=\scriptscriptfont2
          \textfont3=\scriptfont3
          \scriptfont3=\scriptscriptfont3
        \renewcommand{\arraystretch}{0.7}
        \begin{array}{l}}{\end{array}}

\newenvironment{scarray}{
          \textfont0=\scriptfont0
          \scriptfont0=\scriptscriptfont0
          \textfont1=\scriptfont1
          \scriptfont1=\scriptscriptfont1
          \textfont2=\scriptfont2
          \scriptfont2=\scriptscriptfont2
          \textfont3=\scriptfont3
          \scriptfont3=\scriptscriptfont3
        \renewcommand{\arraystretch}{0.7}
        \begin{array}{c}}{\end{array}}


\section{Introduction}  \label{sec_intro}

Antiferromagnetic Potts models \cite{Potts_52,Wu_82,Wu_84}
are much less well understood than their ferromagnetic counterparts.
One reason for this is that the behavior depends strongly
on the microscopic lattice structure,
in contrast to the universality typically enjoyed by ferromagnets.
As a result, many basic questions have to be investigated case-by-case:
Is there a phase transition at finite temperature, and if so, of what order?
What is the nature of the low-temperature phase?
If there is a critical point, what are the critical exponents and the
universality classes?
Can these exponents be understood (for two-dimensional models)
in terms of conformal field theory?

One thing is known rigorously \cite{Kotecky_88,Salas_Sokal_97}:
for $q$ large enough (how large depends on the lattice in question),
the antiferromagnetic $q$-state Potts model has
a unique infinite-volume Gibbs measure and
exponential decay of correlations
at all temperatures, {\em including zero temperature}\/:
the system is disordered as a result of the
large ground-state entropy.
However, for smaller values of $q$, phase transitions can and do occur.
One expects that for each lattice ${\cal L}$ there
will be a value $q_c({\cal L})$ such that
\begin{itemize}
   \item[(a)]  For $q > q_c({\cal L})$  the model has exponential decay
       of correlations uniformly at all temperatures,
       including zero temperature.
   \item[(b)]  For $q = q_c({\cal L})$  the model has a critical point
       at zero temperature.
   \item[(c)]  For $q < q_c({\cal L})$  any behavior is possible.
       Often (though not always) the model has a phase transition
       at nonzero temperature, which may be of either first or second order.
\end{itemize}
The problem, for each lattice, is to find $q_c({\cal L})$
and to determine the precise behavior for each $q \le q_c({\cal L})$.

The $q$-state Potts model on a lattice ${\cal L}$ is defined by the Hamiltonian 
\be
{\cal H} = - J \sum_{\langle \, \vec{x} \, \vec{y}\,  \rangle}
               \delta_{\sigma_{\vec{x}},\sigma_{\vec{y}}}
\ee
where the sum 
$\sum_{\langle \, \vec{x} \, \vec{y}\,  \rangle}$ runs over all possible 
nearest-neighbor pairs of lattice sites (each pair counted once), and each 
spin takes values $\sigma_{\vec{x}} \in \{1,\ldots,q\}$.  
The antiferromagnetic case corresponds to $J = -\beta < 0$.
The Potts antiferromagnet on the hexagonal (honeycomb) lattice 
is the simplest case.  
First, we know that the $q=2$ model has a critical point at $e^J=2-\sqrt{3}$  
because the corresponding ferromagnetic model is critical at $e^J=2+\sqrt{3}$,
and the hexagonal lattice is bipartite. Second, the results of 
\cite{Salas_Sokal_97} show that the hexagonal-lattice antiferromagnet exhibits  
exponential decay of correlations at zero temperature for $q\geq4$.
Thus, $q_c(hc)$ should lie somewhere strictly between $q=2$ and $q=4$. 
The behavior at $q=3$ has not yet been analyzed rigorously.

\begin{figure} 
  \epsfxsize=400pt\epsffile{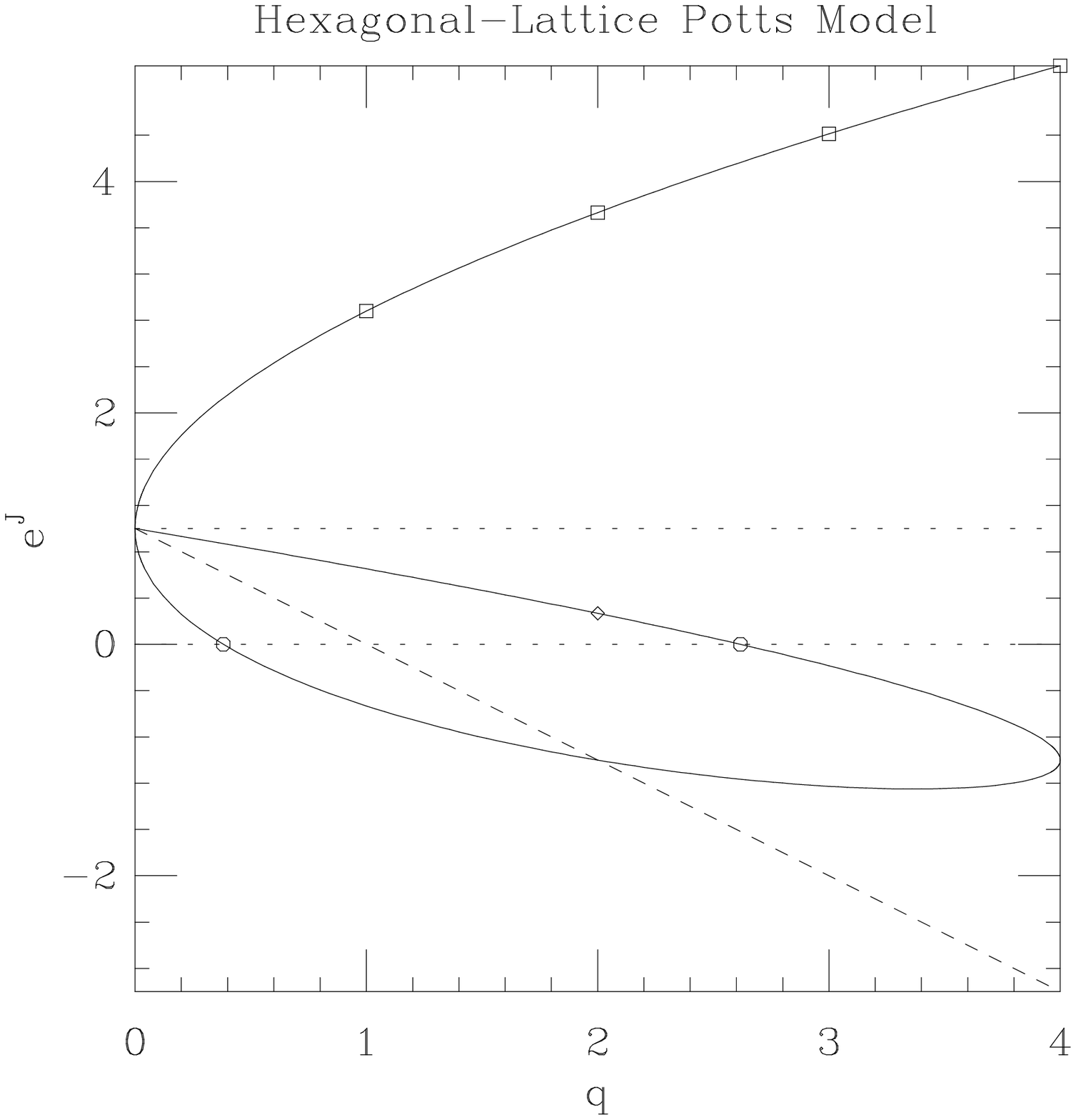}
  \caption[a]{\protect\label{Figure_critical_curves_hexagonal}
  Curves where the hexagonal-lattice Potts model has been solved:
  $(e^J-1)^3 -3q(e^J-1)=q^2$ (solid
  curve), which has three branches; and the line $e^J=1-q$ (dashed line).
  The horizontal dotted lines correspond to $e^J=1$
  (separating the ferromagnetic and antiferromagnetic regimes) and to
  $e^J=0$ (separating the antiferromagnetic regime from the unphysical region
  $e^J<0$).
  The squares ($\Box$) show the known ferromagnetic critical points
  ($q=1,2,3,4$); and the diamond ($\Diamond$) marks the
  known antiferromagnetic critical point for $q=2$.
  The open circles ($\circ$) show the points where
  the two antiferromagnetic branches cross the $e^J=0$ line, namely
  $q = (3 \pm \sqrt{5})/2$.
  }
\end{figure}

On the other hand, the $q$-state hexagonal-lattice Potts model  
has been solved (e.g., the free and internal energies are exactly known) 
on two special curves in the $(J,q)$ plane 
(see Fig.~\ref{Figure_critical_curves_hexagonal}):  
\begin{eqnarray}
\label{hexagonal_curve1}
(e^J - 1)^3 - 3 q (e^J - 1) &\, = \,& q^2  \\ 
\label{hexagonal_curve2}
e^J                   &\, = \,& 1 - q  \qquad \mbox{for $0 < q < 4$}
\end{eqnarray}
These curves are the duality images \cite{Baxter_82a} of  
the corresponding curves for the triangular-lattice Potts antiferromagnet 
obtained by Baxter and collaborators \cite{Baxter_78,Baxter_86,Baxter_87}. 
Curve \reff{hexagonal_curve1} has three branches in the region 
$q\geq0$; the uppermost branch (with $0\leq q\leq \infty$ and 
$e^J \geq 1$) corresponds to the ferromagnetic 
critical line; the middle branch 
(with $0\leq q \leq 4$ and $-1\leq e^J \leq 1$) 
contains the $q=2$ antiferromagnetic
critical point ($e^J = 2 - \sqrt{3}$) and crosses the zero-temperature
line $e^J=0$ at $q = (3 + \sqrt{5})/2 \approx 2.618$; 
the lower-most branch crosses the 
zero-temperature line at $q = (3 - \sqrt{5})/2 \approx 0.382$.\footnote{ 
   The middle branch is missing in Ref.~\protect\cite{Saleur_90}, p. 673, 
   Figure~8. 
}   
The second curve \reff{hexagonal_curve2} is physical ($e^J \geq 0$) 
only for $0\leq q \leq 1$.

The behavior of the middle branch suggest that it may be the 
antiferromagnetic critical curve. If this is the case, there would 
be a zero-temperature critical point at  
\be 
\label{def_qc} 
q = q_c(hc) = { 3 + \sqrt{5} \over 2} 
\ee
(if this assertion has any meaning\footnote{ 
   The Potts models for non-integer $q$ can be given a rigorous meaning
   via the mapping onto the Fortuin-Kasteleyn random-cluster model
   \cite{Fortuin-Kasteleyn_69,Fortuin-Kasteleyn_72,Fortuin_72}.
   The trouble is that in the antiferromagnetic case ($J<0$)
   this latter model has negative weights, and so cannot be given
   a standard probabilistic interpretation.
   In particular, the existence of a good infinite-volume limit
   is problematical;  the limit could depend strongly on the
   subsequence of lattice sizes and on the boundary conditions.
   The same is true of the ``anti-Fortuin-Kasteleyn'' representation,
   in which the coefficients are products of chromatic polynomials of
   clusters: again the weights can be negative for non-integer $q$, and
   the existence of the infinite-volume limit is problematical.
   Likewise, the ice-model representation \cite{Baxter_82a,Baxter_76}
   has in general complex weights for $0<q<4$,
   even in the ferromagnetic case.
})
and the model would be disordered at all temperatures for $q > q_c(hc)$. 
This conclusion is in agreement with the exact result of 
Ref.~\cite{Salas_Sokal_97}. Further interesting speculations can be found
in Ref.~\cite{Saleur_90}.

Shrock and Tsai \cite{Shrock_97a} used this theoretical argument  
to rule out the existence of a critical point in the 3-state antiferromagnet. 
They also pointed out that this argument could fail if there were a 
first-order phase transition (with finite correlation length). 
To test these ideas, they performed Monte Carlo simulations over a 
large range of temperatures ($0 \leq \beta \leq 5$),  
and studied the (static and dynamic) behavior of the 
energy density. They found no signal of hysteresis, and no critical 
slowing-down at large $\beta$ for the Metropolis \cite{Metropolis} and 
Swendsen--Wang cluster \cite{Swendsen_87,Wang_90a} algorithms. 
They finally concluded that 
the model is disordered at all temperatures $T\geq0$.

However, the numerical results of Ref.~\cite{Shrock_97a} 
(namely, smoothness of the energy density and boundedness of its integrated 
autocorrelation time for Metropolis and Swendsen--Wang algorithms) do not
constitute a {\em strong} evidence supporting the absence of criticality of 
this model. A closely related model (e.g., the 3-state square-lattice 
Potts antiferromagnet \cite{Baxter_82a,Baxter_82b}) 
makes an excellent counter-example: it has a critical point at $T=0$, but the  
energy density is smooth over the entire temperature range, and the 
autocorrelation times for the Wang--Swendsen--Koteck\'y (WSK) cluster 
algorithm \cite{Wang_89,Wang_90b}
are uniformly bounded $\tau_{\rm int} \ltapprox 7.6$ 
\cite{Sokal_95,Sokal_98,Ferreira_Sokal_prep}. 
On the other hand, the absence of critical 
slowing-down at large $\beta$ for the (local) Metropolis algorithm only 
implies that the specific heat $C_H$ is bounded. This comes from the  
rigorous bound \cite{Caracciolo_94}:  
\be
  \tau_{{\rm int},{\cal E}} \gtapprox V C_H \; ,
\ee
where $V$ is the volume. Thus, a direct test of non-criticality for the 
3-state hexagonal-lattice Potts antiferromagnet is still lacking. This 
test can be achieved, for instance, by measuring the staggered susceptibility 
and the correlation length.  

Finally, the above-mentioned argument, which identifies exact
solubility with criticality, though plausible, is not necessarily
valid.  For example, in the triangular-lattice case there are
two curves where the model can be solved, and the antiferromagnetic
critical point of the $q=2$ model (namely, $e^J=0$) happens to lie
on both of these curves.  Moreover, the antiferromagnetic critical point
of the $q=4$ model, which is believed 
\cite{Baxter_86,Baxter_70,Henley_96,Henley_prep}\footnote{
   The authors of Ref.~\protect\cite{Henley_96} studied the 3-coloring 
   model on the hexagonal lattice, which is equivalent to the 3-state 
   Potts antiferromagnet on the Kagom\'e lattice {\em at zero temperature}.
   This latter model can be exactly mapped onto the 4-state Potts 
   antiferromagnet on the triangular lattice {\em at zero temperature}
   \protect\cite{Henley_prep}.
}
to be at $e^J=0$,
lies on {\em one}\/ (though not the other) of these curves.
Nevertheless, {\em neither}\/ of these two curves can properly
be identified with the antiferromagnetic critical curve of this model,
as this identification would predict an incorrect scenario for $q=3$
(see Ref.~\cite{Salas_Sokal_97} for details).

In this note, we report the results of performing a direct test of 
non-criticality and we show with no ambiguities that the 3-state Potts 
antiferromagnet on the hexagonal lattice is always disordered. 
In Section~\ref{sec_numerical_simulations} we describe the 
method we have used to simulate the system and the operators we have 
measured. In Section~\ref{sec_data_analysis} we display the results for
the energy, specific heat, staggered susceptibility and second-moment 
correlation length. All these quantities (as well as the corresponding 
integrated autocorrelation times) are bounded {\em uniformly} in 
the temperature and the lattice size. We conclude that the 3-state 
hexagonal-lattice Potts antiferromagnet is disordered at all temperatures
$T\geq0$. As a by-product of our calculation, we compute the zero-temperature
entropy density.

\section{Numerical Simulations} \label{sec_numerical_simulations}

We have performed Monte Carlo simulations of the 3-state hexagonal-lattice 
Potts antiferromagnet at temperatures ranging from $T=\infty$ to $T=0$. 
More precisely, we have simulated this model from $\beta=0$ to 
$\beta=9$ in intervals of 0.05, and also exactly at $\beta=\infty$.  
We have made our simulations using the WSK cluster algorithm. 
This algorithm is suitable to simulate this model {\em at zero temperature} 
because our lattices are bipartite  
\cite{Burton_Henley_97,Ferreira_Sokal_prep}. At $\beta=0$ we have 
started the simulations with a random configuration; at finite $\beta>0$, 
we started from the last configuration generated at the closest smallest
$\beta$; and at $\beta=\infty$ we started from an ordered configuration 
(spins in one sublattice all equal to 1, and spins on the other sublattice
all equal to 2).

The hexagonal lattice is not a Bravais lattice \cite{Ashcroft},  
as not all points are equivalent.
Rather, it is the union of two sublattices, the even and the odd, 
each of which is isomorphic to a triangular lattice (whose lattice spacing
is larger by a factor of $\sqrt{3}$).  
The hexagonal lattice can thus be viewed as an underlying triangular lattice
(which {\em is}\/ Bravais) with a two-point basis.
To be more precise, consider a finite hexagonal lattice ${\cal H}$
with periodic boundary conditions.  Then the even sublattice of
${\cal H}$ is a triangular lattice ${\cal T}$.  Conversely,
given an $L \times L$ triangular lattice ${\cal T}$
with periodic boundary conditions, we can construct a
hexagonal lattice ${\cal H}$ with $V_{hc}=2L^2$ points by taking
the sites in ${\cal T}$ together with the centers of the
down-pointing elementary triangles of ${\cal T}$.  
Thus, a generic point $\vec{x}$ of the hexagonal lattice can be written as 
\cite{Pelissetto_97}:  
\begin{subeqnarray}  
\vec{x} &=& x'_1 \vec{\eta}_1 + x'_2 \vec{\eta}_2 + \epsilon \vec{\eta} 
         \equiv \vec{x}\,' + \epsilon \vec{\eta} \\
x'_1, x'_2 &=& 1,\ldots,L  
\end{subeqnarray}
where $\vec{x}\,'$ lives on the triangular lattice spanned by the 
(unit) vectors
\begin{subeqnarray}   
  \label{def_basis_eta1}
  \vec{\eta}_1 &\; = \;&        (1,0) \\  
  \label{def_basis_eta2}  
  \vec{\eta}_2 &\; = \;&  \left( -{1\over 2},{\sqrt{3}\over 2} \right) \; ,  
\end{subeqnarray}
and $L$ is the linear size of the triangular sublattices.\footnote{ 
   We have chosen the lattice spacing of the triangular sublattice to be 
   $a_{\cal T}=1$; the lattice spacing of the corresponding hexagonal 
   lattice is therefore $a_{\cal H}=1/\sqrt{3}$.
}  
In this paper we have considered lattices ranging from $L=3$ to $L=48$. 
The variable $\epsilon=0,1$ can be interpreted as the ``parity'' of the
corresponding lattice site: if $\epsilon=0$ (resp. $\epsilon=1$) the
site $\vec{x}$ belongs to the even (resp. odd) sublattice. The vector   
\be  
\label{def_basis_eta} 
\vec{\eta}   \; = \;  {1 \over \sqrt{3}} (0,1)  
             \; = \;  {1 \over 3} (\vec{\eta}_1 + 2 \vec{\eta}_2)  
\ee
is the so-called basis vector joining the two points of the basis.  
The pair $(\vec{x}\,',\epsilon)$ determines uniquely 
a point on the hexagonal lattice, and conversely, given  
a point $\vec{x}$ of the hexagonal lattice we can uniquely obtain the 
pair $(\vec{x}\,',\epsilon)$ associated to it.

We have measured three basic observables. The simplest one is the energy 
\be 
\label{def_energy} 
{\cal E} = \sum_{\langle \, \vec{x} \, \vec{y} \, \rangle} 
                 \delta_{\sigma_{\vec{x}} \sigma_{\vec{y}}} \; .  
\ee
The staggered magnetization can be written easily if we represent the 
spin at site $\vec{x}$ by a vector in the plane 
\be 
\vec{\sigma}_{\vec{x}} = \left( \cos {2 \pi \over 3} \sigma_{\vec{x}}, 
                                \sin {2 \pi \over 3} \sigma_{\vec{x}} 
                       \right) \; .  
\ee
In this case, 
the staggering assigns a phase $e^{i\pi\epsilon} = (-1)^\epsilon$ depending
solely on which sublattice the spin is located.\footnote{ 
  This choice is motivated by what happens in the $q=2$ case: at $T=0$ 
  all the spins on the even sublattice take one value (say, 1), and all
  the spins on the other sublattice take the other value (say, 2). The 
  natural staggering corresponds to assign a phase $e^{i \pi}$ to all 
  the spins on the odd sublattice. We can generalize this to the $q=3$ case
  by assigning a general phase $e^{i \phi}$ to all the spins on the 
  odd sublattice. Then, the contribution of the six smallest momenta 
  \protect\reff{smallest_momenta} to the observable \protect\reff{def_F} 
  is the same in average only when $\phi=0$     
 (uniform magnetization), and $\phi=\pi$ (staggered magnetization).
}   
The square of the 
staggered magnetization can be written as  
\be 
\label{def_m2} 
{\cal M}_{stagg}^2 = \left( \sum_{\vec{x}\,',\epsilon} (-1)^{\epsilon} \,  
            \vec{\sigma}_{\vec{x}\,' + \epsilon\vec{\eta}} 
                     \right)^2 = 
                     {3 \over 2} \sum_{\alpha=1}^3 
                    \left| \sum_{\vec{x}\,',\epsilon} 
                    (-1)^\epsilon \,  
                    \delta_{\sigma_{\vec{x}\,'+\epsilon\vec{\eta}},\alpha} 
                    \right|^2 \; .  
\ee
This is a ``zero-momentum'' observable. In order to estimate the 
second-moment correlation length we have to define the corresponding 
``smallest-nonzero-momentum'' observable. 
The translational invariance of the hexagonal lattice ${\cal H}$ is given by
the underlying triangular (Bravais) lattice ${\cal T}$. 
Thus, the allowed momenta are those of a triangular lattice of size 
$L\times L$ with periodic boundary conditions \cite{Ashcroft}:  
\begin{subeqnarray} 
\slabel{allowed_momenta}
\vec{k} &=& {2\pi \over L} m_1 \vec{\rho}_1 + 
            {2\pi \over L} m_2 \vec{\rho}_2 \\
m_1,m_2 &=& 1,\ldots, L 
\end{subeqnarray}  
The momenta are given in the basis 
\begin{subeqnarray}
\vec{\rho}_1 &=& {2 \over \sqrt{3}} \left( {\sqrt{3} \over 2},{1\over 2}
                                  \right) \\
\vec{\rho}_2 &=& {2 \over \sqrt{3}} (0,1) \; , 
\end{subeqnarray}
defined by the relations 
\be
  \vec{\eta}_i \cdot \vec{\rho}_j = \delta_{ij}  \; . 
\ee
The smallest nonzero momenta are 
\be  
 \label{smallest_momenta} 
 \vec{k} = \left\{ \pm {2\pi \over L} \vec{\rho}_1, \; 
                   \pm {2\pi \over L} \vec{\rho}_2, \; 
                   \pm {2\pi \over L} (\vec{\rho}_1 - \vec{\rho}_2) 
           \right\} \; ,  
\ee
all having $|\vec{k}| = 4\pi/(\sqrt{3}L)$. 
Thus, the smallest-nonzero-momentum observable associated to \reff{def_m2} is 
\be  
\label{def_F} 
{\cal F}_{stagg} = {3 \over 2} \times {1 \over 6} 
         \sum_{n=1}^6 
         \sum_{\alpha=1}^3 \left| 
         \sum_{\vec{x}\,',\epsilon} (-1)^\epsilon 
         e^{i \vec{k}_n\cdot(\vec{x}\,' + \epsilon \vec{\eta})}
         \delta_{\sigma_{\vec{x}\,' + \epsilon\vec{\eta}},\alpha} 
                           \right|^2 \; . 
\ee 
The contributions of the wavevectors $\vec{k}_n$ and $-\vec{k}_n$ 
are exactly the same; thus Eq.~\reff{def_F} can be simplified:  
\begin{eqnarray}
{\cal F}_{stagg} = {3 \over 2} \times {1 \over 3} 
         \sum_{\alpha=1}^3  &  &  \left\{ \phantom{+} \;   
         \left|
         \sum_{\vec{x}\,'} e^{2\pi i x'_1/L} \left[  
                            \delta_{\sigma_{\vec{x}\,'},\alpha}  
          -  e^{2 \pi i/3L} \delta_{\sigma_{\vec{x}\,'+\vec{\eta}},\alpha} 
                             \right] \right|^2 \right. \nonumber  \\  
          &   & \quad +  
         \left|
         \sum_{\vec{x}\,'} e^{2\pi i x'_2/L} \left[
                            \delta_{\sigma_{\vec{x}\,'},\alpha}
          -  e^{4 \pi i/3L} \delta_{\sigma_{\vec{x}\,'+\vec{\eta}},\alpha}
                            \right] \right|^2 \nonumber \\ 
          &   &\quad +  
         \left. 
         \left|
         \sum_{\vec{x}\,'} e^{2\pi i (x'_1-x'_2)/L} \left[
                            \delta_{\sigma_{\vec{x}\,'},\alpha}
          -  e^{-2 \pi i/3L} \delta_{\sigma_{\vec{x}\,'+\vec{\eta}},\alpha}
                             \right] \right|^2 
          \right\} 
\label{def_Fbis}
\end{eqnarray} 

From these observables we have computed the following expectation 
values: the energy density (per spin) 
\be 
E = {1 \over V_{hc}} \; \langle {\cal E} \rangle \; ,  
\ee
the specific heat 
\be 
C_H =     {1 \over V_{hc}} \; \mbox{var}({\cal E}) 
   \equiv {1 \over V_{hc}} \; [ \langle {\cal E}^2 \rangle - 
                            \langle {\cal E}\rangle^2 ]  \; , 
\ee
the staggered susceptibility 
\be 
\chi_{stagg} = {1 \over V_{hc}} \langle {\cal M}_{stagg}^2 \rangle \; , 
\ee
and the second-moment correlation length 
\be 
\xi = { (\chi_{stagg}/F_{stagg} - 1)^{1/2} \over 2 \pi \sin(\pi/L)} \; , 
\ee
where $F_{stagg}$ is given by 
\be 
F_{stagg} = {1 \over V_{hc}} \langle {\cal F}_{stagg} \rangle \; . 
\ee

For each observable ${\cal O}$ discussed above we have measured its 
integrated autocorrelation time $\tau_{{\rm int},{\cal O}}$ using 
a self-consistent truncation window of width $6 \tau_{\rm int}$ 
\cite[Appendix C]{Madras_Sokal_88}.

We have made $5 \times 10^5$ (resp $3.5 \times 10^5$) iterations 
for $L=3,6$ (resp. $L\geq 9$). 
We have discarded a 10\% of them to allow the system to reach equilibrium.  
The autocorrelation times for all the observables were uniformly  
bounded in $\beta$ and $L$: 
\be 
  \tau_{\rm int} \ltapprox 4 \; .  
\ee
For $L\geq 9$ there is a sharper bound: $\tau_{\rm int} \ltapprox 3$.\footnote{ 
  The fact that $\tau_{\rm int}$ is larger for smaller lattices can be 
  understood because the correlation length satisfies $\xi \ltapprox 3$ 
  for all $T$ and $L$. For small $L$, the ratio $\xi/L$ is not much smaller
  than 1; however, for large $L$, $\xi/L \ll 1$.  
} 
This means that we have made $\approx 10^5 \, \tau_{\rm int}$ 
measurements, and we have discarded $\approx 10^4 \, \tau_{\rm int}$ 
iterations as a (conservative) prevention against the existence of 
any slower mode not considered here.    

We have made our simulations on two Pentium machines at 166 MHz. Each
WSK iteration took approximately $3 \times (2L^2)$ $\mu$sec; 
the total CPU used was approximately 13 days.

\section{Data Analysis}   \label{sec_data_analysis}

In this section we 
perform all fits using the standard weighted least-squares method.
As a precaution against corrections to scaling,
we impose a lower cutoff $L \ge L_{min}$
on the data points admitted in the fit,
and we study systematically the effects of varying $L_{min}$
on both the estimated parameters and the $\chi^2$.
In general, our preferred fit corresponds to the smallest $L_{min}$
for which the goodness of fit is reasonable
(e.g., the confidence level\footnote{
   ``Confidence level'' is the probability that $\chi^2$ would
   exceed the observed value, assuming that the underlying statistical
   model is correct.  An unusually low confidence level
   (e.g., less than 5\%) thus suggests that the underlying statistical model
   is {\em incorrect}\/ --- the most likely cause of which would be
   corrections to scaling.
}
is $\gtapprox$ 10--20\%)
and for which subsequent increases in $L_{min}$ do not cause the
$\chi^2$ to drop vastly more than one unit per degree of freedom.

\begin{figure}
  \epsfxsize=400pt\epsffile{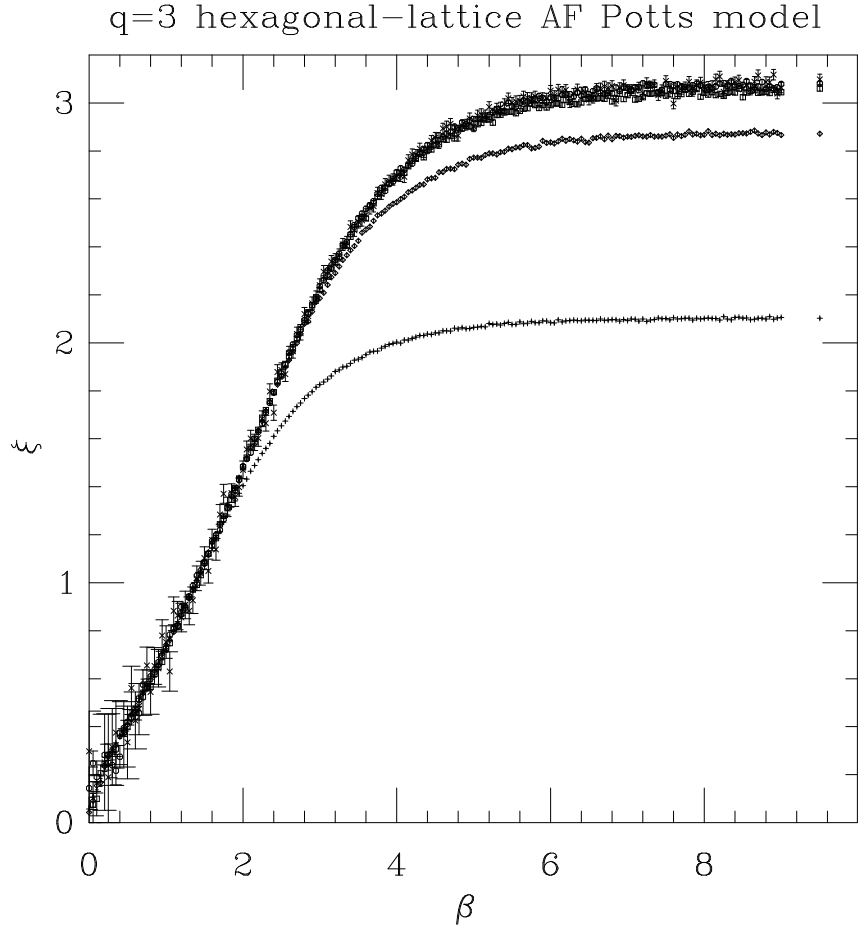}
  \caption[a]{\protect\label{Figure_xi}Second-moment correlation length
  $\xi$ of the
  3-state hexagonal-lattice Potts antiferromagnet as a function of
  $\beta$ and $L$. Symbols indicate $L=3$ ($+$), $L=6$ ($\Diamond$),  
  $L=9$ ($\Box$), $L=12$ ($\circ$), and $L=24$ ($\times$). 
  Points with $L=48$ coincide with $L=24$, 
  and they are not shown for clarity. The isolated points on the far right 
  of the picture (displayed for convenience at $\beta=9.5$) 
  are the data from the runs at $\beta=\infty$. 
  }
\end{figure}

Let us first consider the second-moment correlation length $\xi=\xi(\beta,L)$ 
(See Figure~\ref{Figure_xi}). We see that this observable is, for fixed $L$, 
a non-decreasing function of $\beta$ which asymptotically tends to 
the value at $\beta=\infty$ [i.e., $\xi(\beta=\infty,L)$]. At fixed $\beta$, 
the function $\xi(\beta,\cdot)$ is also non-decreasing. For $L\geq12$,  
the values of $\xi(\beta,L)$ collapses well onto a single curve. 
Furthermore, $\xi(\beta,L) \ltapprox 3.2$ uniformly in $\beta$ and $L$.  
Thus, the correlation length stays bounded even at $T=0$; this observation
implies that there is no critical point for this model at any temperature
$T\geq0$.

If we fit the values of $\xi(\infty,L)$ to a constant [=$\xi(\infty,\infty)$] 
we obtain a good fit for $L_{min}=12$, 
\be 
   \xi(\infty,\infty) = 3.0828 \pm 0.0098 
\ee 
with $\chi^2=1.61$ (2 DF, level = 45\%). Thus, 
$\xi(\beta,L) \leq \xi(\infty,\infty)$  
uniformly in $\beta$ and $L$. The numerical value of 
$\xi(\infty,\infty)$ is consistent with the observation that 
the thermodynamic limit is attained in practice (i.e. 
$\xi(\beta,L) \ll L$) for $L\geq12$.  
Therefore, we do not have to consider larger lattices (i.e., $L>48$).

\begin{figure}
  \epsfxsize=400pt\epsffile{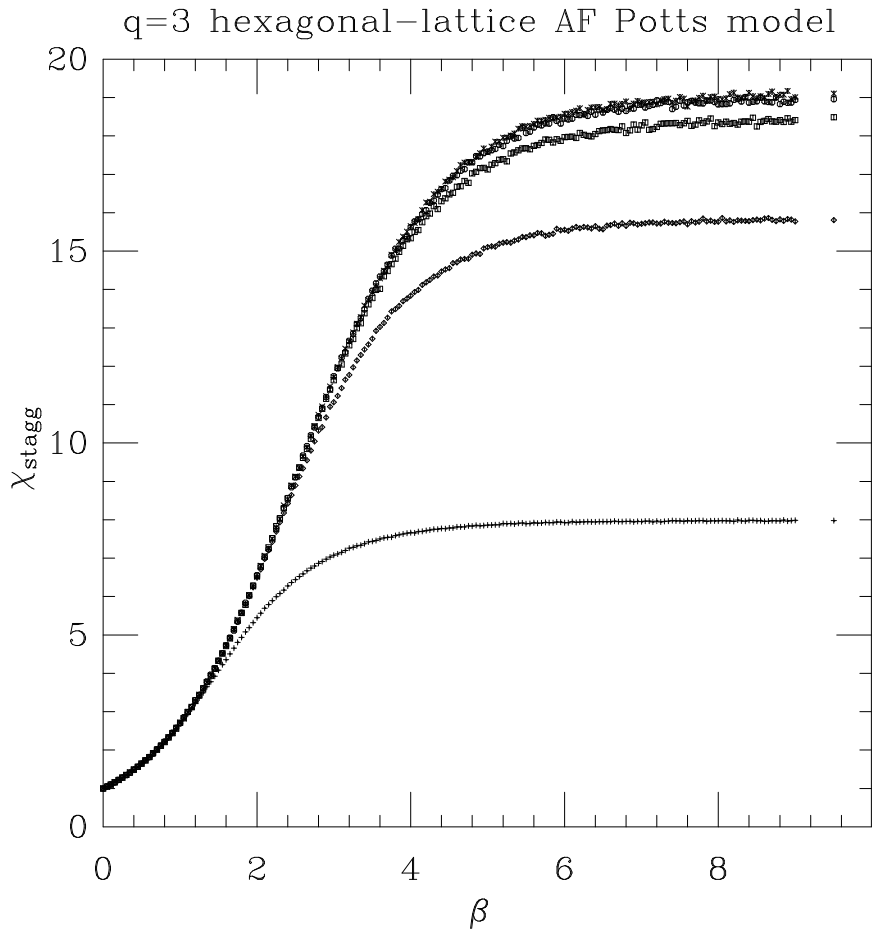}
  \caption[a]{\protect\label{Figure_m2}Staggered susceptibility 
  $\chi_{stagg}$ of the 
  3-state hexagonal-lattice Potts antiferromagnet as a function of 
  $\beta$ and $L$. Symbols are as in Figure~\protect\ref{Figure_xi}.  
  }
\end{figure}

This scenario also applies to the staggered susceptibility (see 
Figure~\ref{Figure_m2}). This observable [$\chi_{stagg}(\beta,L)$] is also  
a non-decreasing function of $\beta$ at fixed $L$, and of $L$ at 
fixed $\beta$. Moreover, for $L\geq 12$ the measurements collapse well 
onto a single curve, which is uniformly bounded: 
$\chi_{stagg}(\beta,L) \ltapprox 20$. There is no signal of second-order or 
first-order phase transition at any temperature.

If we fit the value of the staggered susceptibility at zero temperature 
$\chi_{stagg}(\beta=\infty,L)$ to a constant [$= \chi_{stagg}(\infty,\infty)$],
we obtain a good fit for $L_{min}=24$
\be 
\chi_{stagg}(\infty,\infty)=  19.070 \pm  0.043 
\ee
with $\chi^2=0.48$ (1 DF, level = 49\%).

The energy and specific heat are both non-increasing curves which tend 
asymptotically to zero as $\beta \rightarrow \infty$. For $L \geq 6$ 
the points fall very approximately onto a single curve; for both observables 
the finite-size corrections are very small. Our curve for the energy
coincides with that of Shrock and Tsai \cite{Shrock_97a}. 
The specific heat does not show any signal of transition points: it also
decays smoothly to zero as $\beta\rightarrow\infty$.

In conclusion, there is no signal of phase transitions at any temperature
$T\geq0$ in the 3-state hexagonal-lattice Potts antiferromagnet. This 
model is disordered at all temperatures.

{}From the energy density one can easily compute the entropy density 
(per spin) of this model \cite{Binder_81}:
\be
 \label{def_entropy} 
 S(\beta) \equiv S(0) + \beta E(\beta) - \int_{0}^{\beta} E(\beta') \, d\beta' 
 \; , 
\ee
where the value of the entropy at $\beta=0$ is given by 
$S_{hc,q=3}(\beta=0) = \log q = \log 3$.

Using our numerical data, we are able to compute the value of the entropy 
at $\beta=9$, and we have to extrapolate this value 
somehow to $\beta=\infty$. One way
to achieve this is the following: at very large $\beta$ we are deep in the 
strong coupling limit, so we expect that the energy density behaves as 
$E(\beta) \sim A e^{-\beta} + {\cal O}(e^{-2\beta})$.\footnote{  
   Although there is no obvious way to perform low-temperature series 
   expansion for this model --there are too many inequivalent 
   ground states--, it is 
   reasonable to expect that there is an expansion in powers of 
   $e^{-\beta}$, which corresponds to the minimum energy cost for a 
   ``overturned'' spin. Indeed, our numerical data behaves in this way 
   for large enough $\beta$.  
} 
If this is the case, we can compute exactly the 
integral on the r.h.s. of \reff{def_entropy} and relate the result to 
the energy density at $\beta$:  
\be
 \label{def_entropy_infty}
 S(\infty) = S(0) - E(\beta) - \int_{0}^{\beta} E(\beta') \, d\beta' \; . 
\ee
Now the maximum value of $\beta$ where we have computed the energy becomes
a cutoff. 
For each lattice size $L$, 
we have computed the values of $S(\beta=\infty,L)$ with this method 
using different values of the cutoff $\beta$; the results were consistent 
within statistical errors
(usually the differences were much smaller than the statistical errors).  
The values of the entropy density at zero temperature are displayed 
in Table~\ref{Table_entropy}. The error bars are the sum of the statistical
errors (coming from the statistical errors of the energies) and the 
systematic errors coming from the integration algorithm. We have used 
several extended trapezoidal rules and different sizes of the integration 
intervals \cite{Recipes}: the systematic error takes account 
(conservatively) of the differences we found.

If we fit the data to a constant [$= S_{hc,3}(\infty,\infty)$] 
we find a good fit only for $L_{min}=24$ 
\be 
\label{entropy_res1} 
S_{hc,3}(\infty,\infty) = 0.506844  \pm  0.000012 
\ee
with $\chi^2=0.11$ (1 DF, level = 74\%). This number is in agreement with 
Shrock and Tsai \cite{Shrock_97a,Shrock_97b}: 
\be
\label{entropy_ST}
S_{hc,3}(\infty,\infty)_{ST} = 0.5068  \pm  0.0003 \; , 
\ee
but the error bar is one order of magnitude smaller than in 
Ref.~\cite{Shrock_97b}.
If we use the extended Ansatz of Ref.~\cite{Shrock_97b},   
\be 
\label{ansatz_entropy} 
S_{hc,3}(\beta=\infty,L) = S_{hc,3}(\infty,\infty) + {c_1 \over L^2} + 
  {c_2 \over L^4} + {c_3 \over L^6} \; ,  
\ee
we can reasonably fit all data (i.e., $L_{min}=3$) giving 
\be
\label{entropy_res2}
S_{hc,3}(\infty,\infty) = 0.506841 \pm 0.000018 \; ,  
\ee
with $\chi^2=0.40$ (2 DF, level = 82\%). This estimator agrees within 
errors with \reff{entropy_res1}/\reff{entropy_ST}.\footnote{ 
  An Ansatz of the form 
  $S_{hc,3}(\beta=\infty,L) = S_{hc,3}(\infty,\infty) + c_1/L^2$ is unable
  to fit well the data for any value of $L_{min}$. If we add a
  term $c_2/L^4$, we get a good fit for $L_{min}=6$, giving 
  $S_{hc,3}(\infty,\infty) = 0.506847 \pm 0.000013$ with $\chi^2=0.075$ 
  (1 DF, level = 78\%). The Ansatz \protect\reff{ansatz_entropy} is the 
  first one to be able to fit all the data ($L_{min}=3$). Adding a term 
  $c_4/L^8$ does not modify the conclusions. 
}   

Indeed, the estimator \reff{entropy_res2} is contained within the 
rigorous lower and upper bounds obtained by Shrock and Tsai for 
the hexagonal lattice \cite{Shrock_97c,Shrock_97d}:
\be
\label{bounds} 
{(q^4-5q^3+10q^2-10q+5)^{1/2} \over q-1} \leq e^{S_{hc,q}(\infty,\infty)} 
\leq (q^2 - 3q+3)^{1/2} \; .  
\ee 
For $q=3$ these bounds become 
\be 
0.505800\ldots = \log\left({\sqrt{11}\over 2}\right) \leq 
                 S_{hc,3}(\infty,\infty)             \leq 
\log(\sqrt{3}) = 0.549306\ldots \; .
\ee
The lower bound is remarkably sharp: in Ref.~\cite{Shrock_97d} it is shown 
that if we extract the leading term [$=q (1 -1/q)^{3/2}$] and 
expand the rest in powers of $y=1/(q-1)$, the resulting series for 
$e^{S_{hc,q}(\infty,\infty)}$ and its rigorous lower bound 
[cf. \reff{bounds}] 
agree up to  ${\cal O}(y^{10})$. The lower bound gives a very good 
approximation even for $q$ as low as $q=3$.

The zero-temperature entropy density \reff{entropy_res2} is a large 
fraction ($\approx 46\%$) of the entropy at $T=\infty$ 
[$S_{hc,3}(\beta=0) = \log 3 = 1.09861\ldots$]. 
This large ground-state entropy makes the system 
disordered at zero temperature.

%
%
\section*{Acknowledgments}

We wish to thank Alan Sokal for encouragement and illuminating discussions, 
Robert Shrock for correspondence, 
and Chris Henley for making available some of his unpublished notes.  
The author's research was supported in part by
CICyT grants PB95-0797 and AEN97-1680.

\newpage
\renewcommand{\baselinestretch}{1}
\large\normalsize
%
%
%
%

\clearpage

\newpage
%
%
\begin{table}[htb]
\centering
\begin{tabular}{|r|c|}
\hline \hline 
$L$ & $S_{hc,3}(\infty,L)$ \\
\hline \hline
 3& $0.535387 \pm 0.000158$ \\ 
 6& $0.509207 \pm 0.000097$ \\
 9& $0.507220 \pm 0.000089$ \\
12& $0.506952 \pm 0.000043$ \\ 
24& $0.506864 \pm 0.000062$ \\ 
48& $0.506843 \pm 0.000012$ \\ 
\hline\hline
\end{tabular}
\caption{Values of the 3-state hexagonal-lattice Potts antiferromagnet 
entropy density at zero temperature $S_{hc,3}(\beta=\infty,L)$ 
as a function of the lattice size $L$. The error bars are the sum of 
the statistical and systematic errors.  
}
\label{Table_entropy}
\end{table}


\newpage

%
%

\newpage
%
%

\newpage
%
%

\end{document}